\documentclass[aps, prd, preprintnumbers, twocolumn, amsmath, amssymb, floatfix, nofootinbib, longbibliography]{revtex4-2}
\usepackage[T1]{fontenc}
\usepackage{fancyhdr}
\usepackage[us,12hr]{datetime}
\usepackage{graphicx}
\usepackage{amssymb}
\usepackage{mathrsfs}
\usepackage{marvosym}
\usepackage{caption}
\usepackage{float}
\newcommand{\comment}[1]{}
\usepackage{booktabs}
\usepackage{amsmath,amssymb,amsbsy,amstext, amsthm, simplewick}
\usepackage{graphicx}
\usepackage{ragged2e}
\usepackage[framemethod=tikz]{mdframed}

\newmdenv[
    tikzsetting={draw=black, line width=0.8pt},
    leftmargin=0cm,
    rightmargin=0cm,
    innertopmargin=10pt,
    innerbottommargin=10pt,
    skipabove=10pt,
    skipbelow=10pt
]{mybox}

\usepackage[english]{babel}
\usepackage[utf8]{inputenc}
\usepackage{graphicx}
\usepackage{color}
\usepackage{xcolor}
\usepackage{mathrsfs}
\usepackage{amsmath} % For the equation environment
\usepackage{empheq}  % For boxing the equation
\usepackage[colorlinks=true
,urlcolor=mn%darkcerulean
,anchorcolor=black
,citecolor=magenta
,filecolor=navyblue
,linkcolor=rindou2
,menucolor=navyblue
,pagecolor=navyblue
,linktocpage=true
,pdfproducer=medialab]{hyperref}
\usepackage{slashed}

\usepackage{aas_macros,empheq,fancybox,graphicx,multirow}

\definecolor{pc1}{rgb}{0.69, 0.25, 0.21}

\def\KE{{\sf K}}

\def\GHF{generalized hypergeometric function } 

\newmuskip\pFqmuskip
\newcommand*\pFq[6][8]{
  \begingroup 
  \pFqmuskip=#1mu\relax
  \mathcode`\,=\string"8000
  \begingroup\lccode`\~=`\,
  \lowercase{\endgroup\let~}\pFqcomma
  {}_{#2}F_{#3}{\left[\genfrac..{0pt}{}{#4}{#5};#6\right]}
  \endgroup
}
\newcommand{\pFqcomma}{\mskip\pFqmuskip}

\usepackage{braket}
\usepackage{dsfont}
\usepackage{nccmath}
\usepackage{scalerel}
\usepackage{xcolor}
\definecolor{rindou1}{rgb}{0.4431,0.2862,0.7960}
\definecolor{rindou2}{rgb}{0.0078,0.2215,0.3692}
\definecolor{lapis}{rgb}{0.0270,0.2941,0.5568}
\definecolor{mn}{rgb}{0.15, 0.35, 0.95}
\usepackage{titlesec} 

\begin{document}
\title{\Large \textcolor{rindou2}{On Ising model in magnetic field on the lattice}}
\author{Raghav~G.~Jha}
\email{raghav.govind.jha@gmail.com}
\affiliation{Theory Center, Thomas Jefferson National Accelerator Facility, Newport News, Virginia 23606, United States of America}

\begin{abstract}
\rule{14cm}{0.4pt} \\
\emph{Abstract:}~We study the free energy of the classical square-lattice Ising model in a uniform real magnetic field $h$,
for which no exact solution is known. We first conjecture a closed-form expression built from the
one-dimensional Ising chain and Onsager's hypergeometric solution.
Comparing with high-precision tensor renormalization group (TRG) data, we find that it deviates by up to $2.6\%$,
has no spontaneous magnetization, and gives an effective critical-isotherm exponent $\delta \to 1$.
However, using the Lee--Yang circle theorem, we show that the exact free energy is a superposition of
one-dimensional Ising chains in the same field, weighted by a temperature-dependent (normalized, but not always positive)
distribution of squared Lee--Yang gaps, of which our first expression is a single-chain approximation.
Exact results (Onsager, Yang, the strong-field expansion and the Lee--Yang solution at $\beta h = i\pi/2$)
become sum rules for this distribution. We propose a simple form for it, fixed by Onsager's free energy and the
exact strong-field limit without any fit to numerical data.
This \emph{improved} conjecture agrees with TRG to $0.15\%$ for all $\beta$ and $h$,
reproduces Yang's spontaneous magnetization exactly, has a positive specific heat,
and yields the critical isotherm $M\propto h^{1/\delta}$ with $\delta = 15.2$ and amplitude $1.055$,
compared with the exact $\delta=15$ and $1.059$.
A refined form, which builds in additional input from Ising field theory (including the exact critical isotherm, $\delta=15$)
and a few constants adjusted to transfer-matrix and TRG data, matches the TRG free energy to $4\times 10^{-5}$.
\rule{14cm}{0.4pt} \\
\end{abstract}
\maketitle
\vspace{40mm}
\newpage

%%%%%%%%%%%%%%%%%%%%%%%%%%%%%%%%%%%%%%%
\section{Introduction}
The solution of the two-dimensional Ising model (without magnetic field)
was accomplished by Onsager in his famous work \cite{PhysRev.65.117}.
This marked a significant milestone,
as it was the first instance in which a spin model admitting a phase transition was
solved exactly. Onsager computed the free energy on a square lattice in the absence of a magnetic field using complicated algebraic methods.
Subsequently, Kac and Ward~\cite{PhysRev.88.1332} solved it
using combinatorial methods by expressing the partition
function as a determinant that could be evaluated, and Yang computed the spontaneous magnetization~\cite{Yang1952}.

The two-dimensional model is solvable \emph{only} in the absence of a magnetic field (and at the special imaginary field
$\beta h = i\pi/2$~\cite{PhysRev.87.410, McCoyWu1967, Merlini74}),
and no exact closed-form solution exists in the presence of a non-zero real magnetic field on a
\emph{regular} lattice. In this paper, we revisit this famous unsolved problem in physics. We will focus exclusively on the square lattice.
A great deal is nevertheless known. In the scaling limit ($T \to T_c$ and $h \to 0$) the model is described by the Ising field
theory, which at $T=T_c$ is integrable~\cite{Zamolodchikov:1989fp, Zamolodchikov}; the vacuum energy, and hence the critical
isotherm, is known exactly~\cite{Fateev1994}, and the scaling function of the free energy has been determined to high
precision~\cite{FonsecaZamolodchikov2003, Delfino2004, XuZamolodchikov2022}. On the lattice, corner transfer matrix computations
have confirmed scaling and universality with very high accuracy~\cite{Mangazeev2010, Mangazeev2023}; the zero-field
correlation functions are known exactly and the zero-field susceptibility to arbitrary precision, with exact critical amplitudes~\cite{Wu1966, McCoyWuBook, WMTB1976, Orrick2001}. In the absence of exact
solutions for the two-dimensional model in an external field on the lattice,
several variants of the Ising model that can be exactly solved have been explored. A famous work was done by Berlin and Kac~\cite{PhysRev.86.821}
where they studied a model similar to the Ising model,
known as the `spherical model' based on ferromagnetic nearest-neighbor interactions.
They solved the model in the presence of an external magnetic field
in three dimensions. In the same paper, they conjectured that the solution of the three-dimensional cubic Ising model
can be written as an extended
form of the solution for the two-dimensional case in zero field.
However, this conjecture turned out to be incorrect.\footnote{The conjecture was well-motivated noting the similarity between the functional form of
free energy for the one- and two-dimensional
Ising models: see Ch. 5 of Statistical Mechanics: A Set of Lectures by R.P. Feynman for further details.
}
Soon after, Fisher \cite{Fisher1960} studied a superexchange antiferromagnetic lattice model in which the partition function in the
presence of an external field was computed. This model is closely related to the Ising model,
and is solvable when the field acts on the decorating spins.

In the \underline{first} version of this paper, we conjectured a closed-form expression for the free energy of the Ising model
on a square lattice in a real magnetic field, built from the solution of the one-dimensional Ising chain and Onsager's solution
written in terms of hypergeometric functions~\cite{PhysRevE.83.051101, viswanathan2014hypergeometric}, and compared it with
tensor renormalization group (TRG) computations. Here we revisit that conjecture with considerably more precise numerics
and with the exact information that is available. We find that the original expression [Conjecture~I, Eq.~\eqref{eq:New1} below] is
accurate to $2.6\%$ but misses an essential piece of physics: it has no spontaneous magnetization and its critical isotherm behaves
as $M\sim h\ln(1/h)$ instead of $M \sim h^{1/15}$. We then show, using the Lee--Yang circle theorem~\cite{YangLee1952, PhysRev.87.410},
that the exact free energy can always be written as a superposition of \emph{one-dimensional Ising chains in the same magnetic field},
with a temperature-dependent distribution of (Lee--Yang) gaps. The original conjecture corresponds to a single chain whose gap is fixed by the
energy of a single flipped spin. This suggests a natural improvement, Conjecture~II, Eq.~\eqref{eq:New2}, in which the single chain is replaced by a
distribution with the same mean squared gap, an additional infinitely-coupled chain that carries Yang's spontaneous magnetization, and a shape fixed by
Onsager's free energy; the distribution is constructed such that the non-analytic temperature dependence of the spontaneous magnetization does not leak into
the free energy at $h\neq0$. No parameter is fitted to the numerical data.
The improved expression reduces the maximum deviation from the TRG results by a factor of about twenty (to $0.15\%$) and the root-mean-square deviation by a
factor of thirty, reproduces the spontaneous magnetization exactly, and allows us to \emph{compute} the critical exponent $\delta$ in the limit $h\to 0$:
we obtain $\delta = 15.2$, within $1.5\%$ of the exact value $\delta=15$, and the amplitude of the critical isotherm to $0.4\%$.
We also show that adding further information to the same representation leads to a refined form, Conjecture~II$'$ [Appendix~\ref{app:refined}, Eqs.~\eqref{eq:refined_f}--\eqref{eq:refined_edge}], that agrees with TRG to $4\times10^{-5}$.
The paper is organized as follows. Section~\ref{sec2} collects the exact results at $h=0$ and at $\beta h=i\pi/2$ and describes our numerics; Sec.~\ref{sec:conjI} analyses Conjecture~I;
Sec.~\ref{sec:exact} lists the exact results in a field; Sec.~\ref{sec:chains} derives the chain representation, Eqs.~\eqref{eq:chains}--\eqref{eq:sumrules}; Sec.~\ref{sec:conjII}
presents Conjecture~II and the value of $\delta$; Sec.~\ref{sec:compare} compares the three conjectures with TRG.
The results contained in this paper are neither \emph{exact} nor have a \emph{proof},
but we hope they represent a step in the direction of a solution to this long-standing problem.

\vspace{-3mm}
\section{\label{sec2}\texorpdfstring{$d=2$}{d=2} Ising model on square lattice}

The two-dimensional Ising model has been extensively studied on different regular lattices and on random graphs. The first exact solution was given for the square lattice, and then was extended
to other lattices. The partition function in a uniform field is
\begin{equation}
\label{eq:Z}
\mathbb{Z} = \sum_{\{\sigma\}} e^{\beta \sum_{\langle ij \rangle} \sigma_{i} \sigma_{j} + \beta h \sum_{i} \sigma_{i}},
\end{equation}
where the Boltzmann's constant $k_{B}$, spin coupling $J$,
and magnetic moment $\mu$ have all been set to 1, and $\sigma_i=\pm 1$. We write $f = -\beta^{-1} \lim_{N\to\infty} N^{-1}\ln \mathbb{Z}$,
$H \equiv \beta h$, and $M = \langle \sigma_i\rangle = -\partial f/\partial h$. Without loss of generality we take $h\geq 0$.
In this section, we first consider the standard case of zero field.

\subsection{Exact solution for \texorpdfstring{$h = 0$}{h = 0}}
The solution for the two-dimensional Ising model was first given by Onsager
for zero field. We will not present the full solution here but will simply
quote the results from \cite{PhysRev.65.117}.  This model has phase
transition at a critical value (square lattice), $\beta_{c} = \frac{1}{2}\ln(1+\sqrt{2}) \approx 0.44069$
which is obtained by solving the critical curve $\sinh^{2}(2\beta_{c}) = 1 $
corresponding to the self-dual point.
Exact results for both free energy and its derivatives with respect to $\beta$ are well-known
for the square lattice Ising model (see, for example, Eq.~(109C) of \cite{PhysRev.65.117}).
The free energy\footnote{Another equivalent expression for the free energy is,
\begin{align*}
f(\beta) &= -\frac{1}{\beta}\Bigg( \ln 2  + \frac{1}{8\pi^2}
\int _0^{2\pi }\int _0^{2\pi }\ln \Big[\cosh^{2}(2 \beta) \nonumber \\
&- \sinh(2 \beta)\cos (\phi_1) - \sinh(2 \beta)\cos (\phi_2)\Big]d\phi_1 d\phi_2\Bigg)
\end{align*}
}
is given by:
\begin{align}
\label{eq:Onsager1}
f(\beta) &= -\frac{1}{\beta} \Bigg(\ln (2 \cosh (2 \beta )) +
\frac{1}{ 2 \pi ^2} \int _0^{\pi }\int _0^{\pi }\ln  \nonumber \\
& \left(1-\frac{2 \sinh (2 \beta) \cos (\phi_1)
\cos (\phi_2)}{\cosh ^2(2 \beta )}\right)d\phi_1 d\phi_2\Bigg),
\end{align}
while the internal energy is expressed in terms of an elliptic integral
(Eq.~(118A) of \cite{PhysRev.65.117}) as:
\begin{equation}
\label{eq:UOnsager}
U = -\coth (2\beta) \Bigg(1 + \frac{2}{\pi}\Big(2 \tanh^{2}(2\beta) -1\Big)\KE(4\kappa)\Bigg),
\end{equation}
where $\kappa$ is defined as, $\kappa = \sinh (2\beta)/2\cosh^2(2\beta)$ and the elliptic function
$\KE$ is given by\footnote{Note that in $\textsc{Mathematica 13}$, $\texttt{EllipticK}$ is instead defined as:
\begin{equation*}
\KE(z) = \int_{0}^{\frac{\pi}{2}} \frac{1}{\sqrt{1 - z \sin^2 \phi}} d\phi
\end{equation*}}:
\begin{equation}
\KE(z) = \int_{0}^{\frac{\pi}{2}} \frac{1}{\sqrt{1 - z^2 \sin^2 \phi}} d\phi.
\end{equation}
Though, the expression derived by Onsager in Eq.~\eqref{eq:Onsager1}
is reasonably compact,
one desires an alternate expression without any integrals. Such a closed
expression was first published in 2011~\cite{PhysRevE.83.051101} and studied in detail in Ref.~\cite{viswanathan2014hypergeometric}.
The free energy per site
of two-dimensional Ising model was given
in terms of generalized hypergeometric function
:
\begin{equation}
\label{eq:HPQ1}
f = - \frac{1}{\beta} \Bigg(\ln (2 \cosh (2 \beta))- \kappa^{2} ~
\pFq{4}{3}{1,1,\frac{3}{2},\frac{3}{2}}{2,2,2}{16 \kappa ^{2}}\Bigg),
\end{equation}
where the \GHF is defined as:
\begin{equation}
\pFq{p}{q}{\textbf{a}}{\textbf{b}}{z} = \sum_{n=0}^{\infty}
\frac{(a_{1})_{n} \cdots (a_{p})_{n}}{(b_{1})_{n} \cdots (b_{q})_{n} n!} z^{n},
\end{equation}
where $\textbf{a}$ and $\textbf{b}$ contains $p$ and $q$
elements respectively. Here, we have used $(a)_{n} = \Gamma(a+n)/\Gamma(a) =
a (a+1) \cdots (a+n-1) $
which denotes the rising factorial (also known as `Pochhammer symbol').
As required for convergence, we have $ 16 \kappa ^2 \le 1$ for all $\beta \ge 0$
with phase transition at $16 \kappa^2 = 1$ with $\kappa = \sinh (2\beta)/2\cosh^2{2\beta}$.
It was pointed out in \cite{viswanathan2014hypergeometric} that
Onsager and Glasser had already found a closed form in the 1970s\footnote{Note that
Onsager's solution in terms of double integral
could have been expressed without integrals
by noting the
following result:
\begin{align*}
&\frac{-1}{2\pi^2z^2 } \int_{0}^{\pi} \int_{0}^{\pi}
\ln\Big(1 - 4z \cos \phi_1 \cos \phi_2 \Big) d\phi_1d\phi_2 = \\
&\pFq{4}{3}{1,1,\frac{3}{2},\frac{3}{2}}{2,2,2}{16 z^{2}}.
\end{align*}
We did not find this integral in Gradshteyn and Ryzhik.
The reader can consult Ref.~\cite{SE2023} for the proof.  We thank
David H for the response.
} (unpublished; see also Ref.~\cite{Glasser:2023wvm}). In the form proved in Ref.~\cite{viswanathan2014hypergeometric} it reads
\begin{align}
f &= -\frac{1}{\beta} \Bigg(\ln (2 \cosh (2 \beta)) - \frac{1}{2} + \frac{1}{\pi} \mathsf{E}(4\kappa)  \nonumber \\
&+ \kappa^{2} ~ \pFq{4}{3}{\frac{1}{2},1,1,\frac{3}{2}}{2,2,2}{16 \kappa ^{2}}\Bigg),
\end{align}
where $\mathsf{E}(k) = \int_0^{\pi/2}\sqrt{1-k^2\sin^2\phi}\,d\phi$ is the complete elliptic integral of the second kind; it is
equivalent to Eq.~\eqref{eq:HPQ1}.\footnote{The version of this expression given in the first version of this paper was mistranscribed;
the correct form above agrees with Eq.~\eqref{eq:HPQ1} to machine precision.}
In terms of \GHF
$U(\beta$) is given by:
\begin{align}
\label{eq:UHPQ2}
U(\beta) & =
\frac{-1}{16\cosh^{5}(2\beta)} \Bigg[2 \sinh(10\beta) + 4\sinh(2\beta) + \nonumber \\
& 6\sinh(6\beta) - \Bigg\{\Big[28 \sinh(2\beta) -4\sinh(6\beta)\Big] \nonumber \\
&\pFq{3}{2}{1,\frac{3}{2},\frac{3}{2}}{2,2}{16 \kappa ^{2}}\Bigg\}\Bigg].
\end{align}
It is easy to check that Eq.~\eqref{eq:UOnsager} and Eq.~\eqref{eq:UHPQ2} are equivalent (the prefactor $1/16$ was missing in the first version).
Finally, the spontaneous magnetization, $M_0(\beta) = \lim_{h\to 0^+} M(\beta,h)$, was computed by Yang~\cite{Yang1952}:
\begin{equation}
\label{eq:Yang}
M_0(\beta) = \Big(1 - \sinh^{-4}(2\beta)\Big)^{1/8} \quad (\beta>\beta_c),
\end{equation}
and $M_0 = 0$ for $\beta\le\beta_c$. Since $M_0 \ne 0$ for $\beta > \beta_c$, the free energy is \emph{not} an analytic function of $h$ at $h=0$
in the ordered phase: $f(\beta,h) = f(\beta,0) - M_0 |h| + \cdots$, i.e., $h=0$, $T<T_c$ is a line of first-order transitions which ends at the
critical point $(T_c, h=0)$~\cite{Yang1952, PhysRev.87.410, Isakov1984}.

\subsection{Exact solution at the imaginary field \texorpdfstring{$\beta h = i\pi/2$}{beta h = i pi/2}}
There is a second exactly solvable point. At $z \equiv e^{-2\beta h} = -1$, i.e., $\beta h = i\pi/2$, Lee and Yang gave the free energy and the magnetization
(Eqs.~(48) and (49) of Ref.~\cite{PhysRev.87.410}); these were later derived and studied in Refs.~\cite{McCoyWu1967, Merlini74, MatveevShrock1995}.
In the notation used above, the real part of the free energy is
\begin{widetext}
\begin{align}
\label{eq:Merlini1}
-\beta\, {\rm Re}\, f\Big(\beta, \frac{i\pi}{2\beta}\Big) &= \ln 2  + \frac{1}{16\pi^2}
\int _0^{2\pi }\int _0^{2\pi } \ln \Big[\sinh^{2}(2 \beta) \Big( 1 + \sinh^{2}(2 \beta) - \sin\phi_1 \sin\phi_2\Big)
\Big] d\phi_1 d\phi_2 \nonumber \\
&= \frac{1}{2}\ln\big(2\sinh 4\beta\big) - \frac{1}{32\cosh^4(2\beta)}\,
\pFq{4}{3}{1,1,\frac{3}{2},\frac{3}{2}}{2,2,2}{\frac{1}{\cosh^4(2\beta)}},
\end{align}
\end{widetext}
where the second line follows from the same integral identity that leads to Eq.~\eqref{eq:HPQ1}, and the imaginary part of
$-\beta f$ is $\pi/2$. The magnetization at this field (approached from ${\rm Re}\,h>0$) is real and given by the remarkably simple
formula~\cite{PhysRev.87.410, MatveevShrock1995}
\begin{equation}
\label{eq:LYM}
M\Big(\beta, \frac{i\pi}{2\beta}\Big) = \Big(1 - \cosh^{-4}(2\beta)\Big)^{-1/8},
\end{equation}
which should be compared with Yang's result, Eq.~\eqref{eq:Yang}. Eq.~\eqref{eq:LYM} is analytic at $\beta_c$, where it equals $2^{1/4}3^{-1/8}$.

\subsection{Numerical method}
\begin{figure}
\centering
\includegraphics[width=0.45\textwidth]{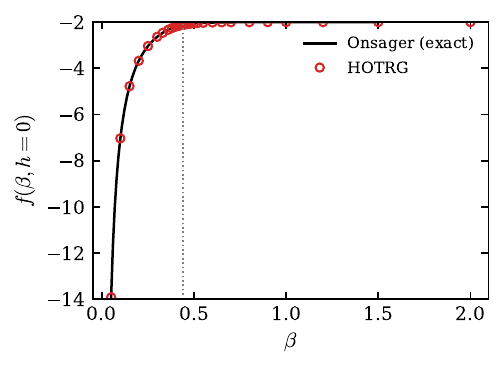}
\caption{\label{fig:1} Comparison of the numerical HOTRG results (hollow circles)
with the exact result of Onsager~\cite{PhysRev.65.117} for $h=0$.}
\end{figure}

Before we proceed to discuss the conjecture for non-zero
real magnetic field, we show that the numerical method we use
to check our conjecture is a robust one.
We use the higher-order tensor renormalization group (HOTRG)~\cite{Xie2012}, a variant of the
real-space tensor renormalization group~\cite{PhysRevLett.99.120601}.
The initial tensor is constructed by expanding the Boltzmann weights,
$e^{\beta \sigma \sigma'} = \sum_a W_{\sigma a} W_{\sigma' a}$ with $W = \big(\begin{smallmatrix} \sqrt{\cosh\beta} & \sqrt{\sinh \beta} \\
\sqrt{\cosh\beta} & -\sqrt{\sinh\beta}\end{smallmatrix}\big)$, and $T_{abcd} = \sum_{\sigma} e^{\beta h\sigma} W_{\sigma a}W_{\sigma b}W_{\sigma c}W_{\sigma d}$,
so that $\mathbb{Z} = {\rm tTr} \prod T$. Coarse-graining alternately in the two directions we reach lattices of
$2^{36}$ sites (thermodynamic limit) while keeping $D$ states; the magnetization is computed with an impurity tensor in which
$e^{\beta h\sigma}$ is replaced by $\sigma e^{\beta h\sigma}$.
One of the advantages of the tensor method is that the thermodynamic
limit is only logarithmically expensive in system size (after fixing the truncation)
unlike Monte Carlo, and that the free energy, which is required to check the conjecture, is the easiest observable to
compute with good accuracy. At $h=0$ the relative error of the free energy, compared with Eq.~\eqref{eq:HPQ1}, is about
$6\times10^{-7}$ at $\beta_c$ for $D=16$, and $1\times 10^{-7}$ and $3\times 10^{-8}$ for $D=24$ and $32$;
for $h\neq 0$ the correlation length is finite and the convergence is faster. The comparison at $h=0$ is shown in Fig.~\ref{fig:1}.
We have computed $f$ and $M$ on a grid of $34$ values of $\beta\in[0.05, 2]$ (dense near $\beta_c$) and $25$ values of $h \in [0, 5]$
with $D=16$, and with $D=24$ for $0.38\le\beta\le0.55$ and nine other temperatures; the two differ by at most $5\times 10^{-7}$ in $f$ (relative),
the largest differences occurring near $\beta_c$ at $h=0$. We have cross-checked the whole grid against an independent corner transfer matrix
renormalization group computation and, at selected points, against exact transfer matrices on cylinders of circumference up to $20$: the free energy
we use is accurate to $1\times10^{-7}$ (relative) everywhere, far below the deviations of the conjectures discussed below.
At $\beta_c$ and with $D=32$, our TRG magnetization agrees with the exact critical isotherm,
Eq.~\eqref{eq:isotherm} below, to $0.22\%$ for $10^{-7}\le h \le 10^{-2}$. The same code with complex $h$, at $\beta h = \epsilon + i\pi/2$ with $\epsilon=0.005$, reproduces Eq.~\eqref{eq:Merlini1} (after subtracting the shift $\epsilon M$)
and Eq.~\eqref{eq:LYM} to about $2\times10^{-4}$ for $\beta\geq 0.3$; at higher temperatures the residual $\epsilon$-dependence of $M$ is larger (about $10^{-3}$ at $\beta=0.2$).

%------------------------------------------------------------------------------------------------
%---------------------------------------------------------------------------------------------

%%%%%%%%%%%%%%%%%%%%%%%%%%%%%%%%%%%%%%%
\section{The v1 conjecture for \texorpdfstring{$h \neq 0$}{h != 0} and its limitations}
\label{sec:conjI}
In the presence of a magnetic field, the $\mathbb{Z}_{2}$ symmetry is broken
and the usual algebraic and combinatorial methods fail. The exact results which are
known correspond to $ z = e^{-2 \beta h}$ equal to $1$ (Onsager)
or $-1$ (Lee--Yang). In this work, we focus on the region
$ 0 \le  z < 1$ where the partition function is not known.
Our original work was partly inspired by Fisher \cite{Fisher1960} where
the partition function in the presence of an external field was computed for a
superexchange antiferromagnetic lattice model, which
is derived from the deformation of the usual square-lattice Ising model
by decorating each bond of the standard lattice with an extra spin.
Although the Ising model with non-zero $h$ is unsolved,
the other model is solvable~\cite{Fisher1960}.
In the first version of this paper we conjectured:
\begin{widetext}
\begin{minipage}{0.98\textwidth}
    \begin{mdframed}[backgroundcolor=blue!0]
\begin{align}
\label{eq:New1}
f_{\rm I}(\beta, h) &= -\frac{1}{\beta} \ln \Bigg( \cosh (\beta h )e^{2\beta} + \sqrt{e^{-4\beta} + \sinh ^2(\beta h)
e^{4\beta}} \Bigg) + \frac{1}{\beta} \xi^{2} ~ \pFq{4}{3}{1,1,\frac{3}{2},\frac{3}{2}}{2,2,2}{16 \xi ^{2}},
\end{align}
where
\begin{equation*}
\xi = \frac{\cosh (\beta h )e^{2\beta} -\sqrt{e^{-4\beta} + \sinh ^2(\beta h)
e^{4\beta}}}{\Big(\cosh (\beta h )e^{2\beta} + \sqrt{e^{-4\beta} + \sinh ^2(\beta h) e^{4\beta}}\Big)^{2}}.
\end{equation*}
     \end{mdframed}
\end{minipage}
\end{widetext}
\textbf{We refer to Eq.~\eqref{eq:New1} as Conjecture~I}. The argument of the logarithm is the largest eigenvalue $\lambda_+$ of the transfer matrix of a one-dimensional Ising chain with
coupling $2\beta$ (two bonds per site) in the field $h$, and $\xi = \lambda_-/\lambda_+^2$ reduces to $\kappa$ at $h=0$, so that
Eq.~\eqref{eq:New1} reduces to Onsager's result, Eq.~\eqref{eq:HPQ1}. The chain term is exact to leading order at strong field or low
temperature, $-\beta f = 2\beta + \beta h + e^{-8\beta - 2\beta h} + \cdots$, which corresponds to a single flipped spin in the ordered background.

\begin{figure}[tbp]
	\centering
	\includegraphics[width=0.45\textwidth]{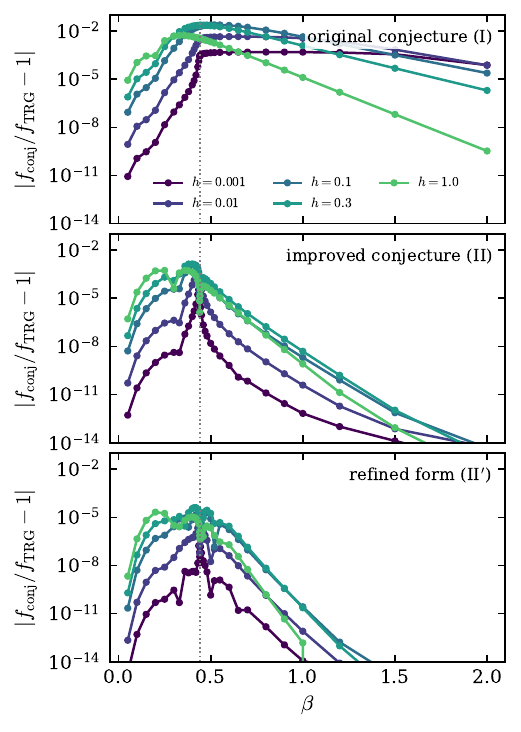}
	\caption{\label{fig:error}Relative deviation of the free energy from HOTRG, $|f_{\rm conj}/f_{\rm TRG}-1|$, for
	(top) Conjecture~I, the original conjecture, Eq.~\eqref{eq:New1};
	(middle) Conjecture~II, the improved conjecture, Eq.~\eqref{eq:New2} with $(a,b)$ fixed by the first two sum rules of Eq.~\eqref{eq:sumrules};
	(bottom) Conjecture~II$'$, the refined form of Appendix~\ref{app:refined}, Eqs.~\eqref{eq:refined_f}--\eqref{eq:refined_edge}.
	The dotted line marks $\beta_c$.}
\end{figure}

\begin{figure}[tbp]
	\centering
	\includegraphics[width=0.45\textwidth]{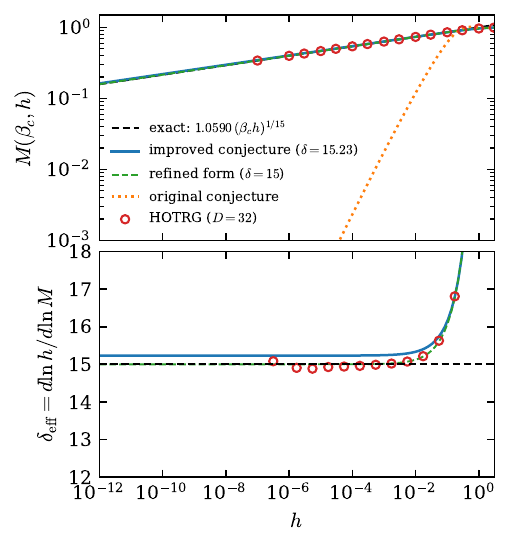}
	\caption{\label{fig:isotherm}Top: magnetization on the critical isotherm from HOTRG with $D=32$ (circles), Conjecture~II [Eq.~\eqref{eq:New2}, i.e.\ Eq.~\eqref{eq:M2F1}] (solid),
	Conjecture~II$'$ [Eq.~\eqref{eq:refined_M}] (green dashed) and Conjecture~I [Eq.~\eqref{eq:New1}] (dotted); the black dashed line is the exact result, Eq.~\eqref{eq:isotherm}.
	Bottom: effective exponent $\delta_{\rm eff} = d\ln h/d\ln M$ (for the original conjecture $\delta_{\rm eff}\to1$, outside the range shown).}
\end{figure}

With the more precise data described above we find that the maximum relative deviation of Eq.~\eqref{eq:New1} from TRG is
$2.6\%$ (at $\beta = 0.5$, $h = 0.15$), with a root-mean-square deviation of $0.9\%$ over our grid (Fig.~\ref{fig:error}, upper panel).
More importantly, the deviation has a clear physical origin. Eq.~\eqref{eq:New1} is an even function of $h$ which is analytic at $h=0$ for all $\beta\neq\beta_c$.
Therefore it has \emph{no} spontaneous magnetization: $M_{\rm I}(\beta, 0^+) = 0$ for all $\beta$, whereas the exact result is
Eq.~\eqref{eq:Yang}. Indeed, for $\beta>\beta_c$ and small $h$, the error of Eq.~\eqref{eq:New1} in $-\beta f$ is precisely $\beta h M_0(\beta)$
(e.g., $5.8\times 10^{-4}$ at $\beta = 0.6$, $h = 10^{-3}$). The statement in the first version that the absence of singularities in
Eq.~\eqref{eq:New1} is consistent with the Lee--Yang and Lebowitz--Penrose theorems~\cite{lebowitz1968} was therefore too quick:
the exact free energy is analytic for real $h\neq 0$, but at $h=0$ it is singular for all $T\le T_c$.
At the critical temperature, $1-16\xi^2 \propto h^2$ and the $\,_4F_3$ term gives $\Delta f\propto h^2 \ln h$, i.e.,
$M_{\rm I}(\beta_c, h)\sim h\ln(1/h)$: the effective critical-isotherm exponent of Eq.~\eqref{eq:New1} tends to $\delta=1$, while the exact value is
$\delta=15$ (see Fig.~\ref{fig:isotherm}). These defects are cured by Conjecture~II, Eq.~\eqref{eq:New2} of Sec.~\ref{sec:conjII}, and by its refined version,
Conjecture~II$'$, Eqs.~\eqref{eq:refined_f}--\eqref{eq:refined_edge} in Appendix~\ref{app:refined}; the deviations of all three from TRG are compared in Fig.~\ref{fig:error}.


\section{Exact results in a magnetic field}
\label{sec:exact}
Any improved expression should be consistent with the following exact results.

\emph{(i) Critical isotherm.} At $T=T_c$, the singular part of the free energy is given by the vacuum energy of Zamolodchikov's $E_8$ integrable
field theory~\cite{Zamolodchikov:1989fp}, computed exactly by Fateev~\cite{Fateev1994},
$-\beta \Delta f = 1.197733383\ldots\, (s\, \beta h)^{16/15}$, where $\Delta f = f(\beta_c, h) - f(\beta_c, 0)$ and $s$ relates the lattice spin to the
continuum field, $s^2 = 2^{5/24} e^{3\zeta'(-1)} = 0.703380\ldots$, the amplitude of the critical spin-spin correlation
function~\cite{Wu1966, McCoyWuBook}. Hence,
\begin{align}
\label{eq:isotherm}
-\beta \Delta f &= 0.992799\, (\beta h)^{16/15}, \nonumber \\ M(\beta_c, h) &= 1.058986\, (\beta h)^{1/15} \, ,
\end{align}
up to corrections of relative order $h^{14/15}$. Our TRG data agree with Eq.~\eqref{eq:isotherm} (Fig.~\ref{fig:isotherm}).
The exponent $\delta=15$ is related to the zero-field exponents $\beta_m = 1/8$ (Yang) and $\gamma=7/4$~\cite{WMTB1976} by $\delta = 1+\gamma/\beta_m$~\cite{Widom1965};
the zero-field susceptibility $\chi = \partial M/\partial(\beta h)|_{h\to0^+}$ behaves as $C_0^{\pm}|t|^{-7/4}$ with $t = 1-T_c/T$, $C_0^+ = 0.9625817$ and
$C_0^- = 0.0255370$~\cite{WMTB1976, Orrick2001}.

\emph{(ii) Strong-field expansion.} Expanding in $z=e^{-2\beta h}$ at fixed $u = e^{-4\beta}$ (clusters of flipped spins in the fully ordered state),
\begin{equation}
\label{eq:strong}
-\beta f = 2\beta + \beta h + \sum_{n\ge 1} L_n(u)\, z^n,
\end{equation}
with $L_1 = u^2$, $L_2 = 2u^3 - \tfrac{5}{2}u^4$, $\ldots$ The polynomials $L_n(u)$ are exact; $L_1$ is the single flipped spin, with energy cost $8\beta + 2\beta h$.

\begin{figure}[tbp]
	\centering
	\includegraphics[width=0.45\textwidth]{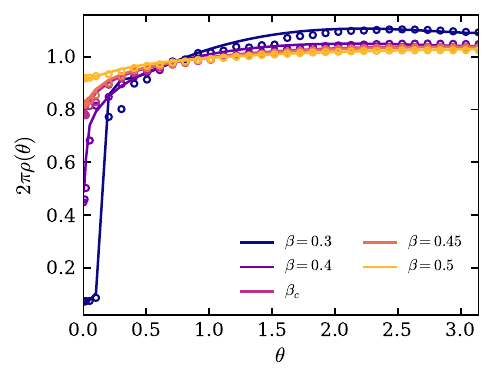}
	\caption{\label{fig:LY}Density of Lee--Yang zeros, $2\pi\rho(\theta) = {\rm Re}\, M$ at $\beta h = \epsilon - i\theta/2$ with $\epsilon=0.01$,
	from TRG at complex field (markers) and from Conjecture~II, Eqs.~\eqref{eq:New2} and \eqref{eq:Mchains} (lines, evaluated at the same $\epsilon$).}
\end{figure}

\emph{(iii) Lee--Yang theorem.} The zeros of $\mathbb{Z}$, viewed as a polynomial in $z$, lie on the unit circle,
$z_j = e^{i\theta_j}$~\cite{PhysRev.87.410}. In the thermodynamic limit they have a density $\rho(\theta) = \rho(-\theta)\ge 0$,
$\int_{-\pi}^{\pi} \rho\, d\theta = 1$, and
\begin{equation}
\label{eq:LYrep}
-\beta f(\beta, h) = 2\beta + \frac{1}{2}\int_{-\pi}^{\pi} d\theta\, \rho(\theta) \ln\Big[ 4\sinh^2(\beta h) + 4\sin^2\frac{\theta}{2}\Big] .
\end{equation}
The density is the real part of the magnetization at imaginary field, $2\pi\rho(\theta) = {\rm Re}\,M(\beta, h)|_{\beta h = 0^+ - i\theta/2}$; in
particular $2\pi\rho(0) = M_0$ and $2\pi\rho(\pi)$ is given by Eq.~\eqref{eq:LYM}. Above $T_c$ the zeros have a gap $|\theta|<\theta_0(\beta)$ which closes
at $T_c$, with the Yang--Lee edge singularity $\rho \propto (\theta-\theta_0)^{-1/6}$~\cite{Fisher1978, Cardy1985}; in the scaling limit
the edge is at $\beta h = i\xi_0 s^{-1} \mu^{15/8}$ with $\xi_0 \simeq 0.1893$ [$0.18930(5)$~\cite{FonsecaZamolodchikov2003}, $0.18933$~\cite{XuZamolodchikov2022}, $0.189351$~\cite{Mangazeev2023}; we use $0.18933$],
where $\mu = -\ln\tanh\beta - 2\beta$ is the exact inverse correlation length above $T_c$~\cite{McCoyWuBook}.
At $T_c$, $\rho(\theta)\propto \theta^{1/\delta}$~\cite{Kim2006}.
We have computed $\rho(\theta)$ by running our TRG at complex $h$ (Fig.~\ref{fig:LY}); see also Ref.~\cite{GarciaSaezWei2015}.

\section{Ising model as a superposition of Ising chains}
\label{sec:chains}
For the one-dimensional Ising chain with coupling $K$ in the field $h$, the largest transfer-matrix eigenvalue is
$\lambda_+ = e^{K}\big[\cosh H + \sqrt{\sinh^2 H + v}\,\big]$ with $v = e^{-4K}$, and Eq.~\eqref{eq:LYrep} holds (with $2\beta\to K$) with
$2\pi\rho = |\sin\frac{\theta}{2}| / \sqrt{\sin^2\frac{\theta}{2} - v}$ for $\sin^2\frac{\theta}{2}>v$ and zero otherwise; the zeros have a gap
$|\theta|<2\arcsin\sqrt{v}$, and we call $v$ the squared Lee--Yang gap.
Since the kernel of Eq.~\eqref{eq:LYrep} depends on $\theta$ only through $\sin^2\frac{\theta}{2}$, the density of any model obeying the Lee--Yang
theorem can be decomposed, by an Abel transformation, into such one-dimensional densities: $2\pi\rho(\theta) = \int_0^{\varsigma^2} dW(v)\, \varsigma/\sqrt{\varsigma^2 - v}$
with $\varsigma = |\sin\frac{\theta}{2}|$, whose inverse is
$W(v) = \frac{2}{\pi}\int_0^{\pi/2} 2\pi \rho\big(2\arcsin(\sqrt{v}\sin\varphi)\big)\, d\varphi$. One thus obtains the \emph{exact} representation
\begin{align}
\label{eq:chains}
-\beta f(\beta, h) = 2\beta &+ \int_0^1 dW_\beta(v) \nonumber \\
&\times \ln\Big[\cosh(\beta h) + \sqrt{\sinh^2(\beta h) + v}\,\Big],
\end{align}
with $\int_0^1 dW_\beta = 1$, so that
\begin{equation}
\label{eq:Mchains}
M(\beta, h) = \int_0^1 dW_\beta(v)\, \frac{\sinh (\beta h)}{\sqrt{\sinh^2 (\beta h) + v}} \, .
\end{equation}
In words: the free energy of the two-dimensional Ising model in a field is a superposition of free energies of one-dimensional Ising chains
in the \emph{same} field, with squared Lee--Yang gaps $v$ (effective couplings $K = -\frac14\ln v$) distributed according to $W_\beta$.
The exact $W_\beta$ is normalized but need not be positive: for a positive $W_\beta$ the magnetization at $\beta h = \epsilon + i\pi/2$ must decrease
with $\epsilon$ [each chain gives $\cosh\epsilon/(\cosh^2\epsilon - v)^{1/2}$], while our TRG data and exact transfer matrices show that it increases for $\beta\lesssim0.25$.
For $\beta\gtrsim 0.3$ the data are consistent with $W_\beta\ge0$.
The leading term of Eq.~\eqref{eq:New1} is the single chain $dW = \delta(v - e^{-8\beta})\,dv$.
The exact information of Sec.~\ref{sec:exact} translates into simple properties of $W_\beta$:
\begin{itemize}
\item[(a)] Yang: $W_\beta$ has an atom of weight $M_0(\beta)$ at $v=0$ (an infinitely coupled chain, which contributes $M_0 |\beta h|$).
\item[(b)] Sum rules. Comparing Eq.~\eqref{eq:chains} with Eqs.~\eqref{eq:HPQ1}, \eqref{eq:Merlini1}, \eqref{eq:LYM} and \eqref{eq:strong} (the expansion of the
chain free energy is $\ln[\cosh H + \sqrt{\sinh^2 H + v}] = H + v z + (v - \frac{3}{2}v^2) z^2 + \cdots$),
\begin{align}
\label{eq:sumrules}
\int dW_\beta \ln(1+\sqrt{v}) &= -\beta f(\beta,0) - 2\beta, \nonumber \\
\int dW_\beta\, v &= e^{-8\beta}, \nonumber \\
\int dW_\beta\, v^2 &= \tfrac{2}{3}u^2 - \tfrac{4}{3}u^3 + \tfrac{5}{3} u^4, \nonumber \\
\int dW_\beta\, \tfrac{1}{2}\ln(1-v) &= -\beta\,{\rm Re} f\big(\beta,\tfrac{i\pi}{2\beta}\big) - 2\beta, \nonumber \\
\int dW_\beta\, (1-v)^{-1/2} &= M\big(\beta,\tfrac{i\pi}{2\beta}\big).
\end{align}
In general the strong-field polynomials $L_n(u)$ fix all integer moments of $W_\beta$. The single chain of Eq.~\eqref{eq:New1} satisfies the second rule exactly,
which is why it is accurate at strong field.
\item[(c)] Critical isotherm: $dW_{\beta_c} \propto v^{1/(2\delta)-1} dv$ as $v\to 0$, so that $M(\beta_c,h)\propto h^{1/\delta}$. Since this distribution carries
a finite fraction ($\propto t^{1/8}$) of its weight at $v\lesssim |t|^{15/4}$, analyticity of $f$ in $\beta$ at fixed $h\ne0$ requires that the changes of $W_\beta$ near $T_c$
be \emph{local}: weight removed from small $v$ must reappear at small $v$ (in the atom of (a) below $T_c$, near the Yang--Lee edge above $T_c$).
\item[(d)] Above $T_c$: $W_\beta$ is supported on $[v_0, 1]$, with $v_0 = \sin^2\frac{\theta_0}{2}$, and $dW \propto (v-v_0)^{-2/3}dv$ near the Yang--Lee edge.
\item[(e)] Limits: at low temperature $dW \to M_0\, \delta(v)\, dv + (1-M_0)\, dv$; at high temperature $W_\beta$ concentrates at a single chain.
\end{itemize}
Any positive measure $W_\beta$ gives a free energy that obeys the Lee--Yang theorem and is therefore analytic for real $h\neq 0$; moreover,
$0\le M \le 1$, and $M$ is increasing and concave in $h>0$ (each chain term is). The conjectures discussed in this paper are particular choices of $W_\beta$ (or of $\rho$):
the single chain of Conjecture~I, Eq.~\eqref{eq:New1}; the distribution of Conjecture~II, Eq.~\eqref{eq:New2}; and the zero density of Conjecture~II$'$,
Eqs.~\eqref{eq:refined_m}--\eqref{eq:refined_edge}.

\section{Improved conjecture}
\label{sec:conjII}
The simplest distributions compatible with (a)--(e) lead to our improved conjecture, \textbf{which we call Conjecture~II. It is defined by Eq.~\eqref{eq:New2} below}, with the
magnetization given by Eq.~\eqref{eq:Mchains} and the two shape parameters $(a,b)$ fixed by the first two sum rules of Eq.~\eqref{eq:sumrules}:
\begin{widetext}
\begin{minipage}{0.98\textwidth}
    \begin{mdframed}[backgroundcolor=blue!0]
\begin{align}
\label{eq:New2}
f_{\rm II}(\beta, h) &= -\frac{1}{\beta}\left\{ 2\beta + \int_0^1 dW_\beta(v)\, \ln\Big[\cosh(\beta h) + \sqrt{\sinh^2(\beta h) + v}\,\Big] \right\}, \\
dW_\beta(v) &= M_0\, \delta(v)\, dv + \frac{(v+v_1)^{a-1}(1-v)^{b-1}}{B(a,b)}\, dv \qquad (\beta\ge\beta_c), \nonumber\\
dW_\beta(v) &= \mathcal{N}^{-1}\, (v-v_0)^{-2/3}\, v^{a-1/3} \Big[1+\omega \Big(\frac{v_0}{v}\Big)^{n}\Big] (1-v)^{b-1}\,\Theta(v-v_0)\, dv \qquad (\beta<\beta_c), \nonumber \\
v_0 &= \tanh^{15/4}\!\big(\ell \mu\big), \qquad \mu = -\ln\tanh\beta - 2\beta, \qquad \ell = (\xi_0/s)^{8/15} = 0.45213, \nonumber \\
\omega &= -\frac{B(-a,\frac13)}{B(n-a,\frac13)}, \qquad n = 1 + 6 v_0. \nonumber
\end{align}
\end{mdframed}
\end{minipage}
\end{widetext}
Here $M_0$ is Yang's result Eq.~\eqref{eq:Yang}, $B(x,y)$ is the Euler beta function, and $\mathcal{N}$ normalizes $W_\beta$. Below $T_c$, $v_1\ge0$ is fixed by the
normalization $\int dW_\beta = 1$; it vanishes at $T_c$ and is of order $[a B(a,b) M_0]^{1/a}\propto |t|^{1/(8a_c)}\approx|t|^{3.8}$ close to it. Above $T_c$, $\ell$ is chosen so that
$v_0^{1/2}$ reproduces the Yang--Lee edge of the Ising field theory as $T\to T_c^{+}$ (Sec.~\ref{sec:exact}); away from $T_c$ this formula agrees with the
edge we measure with complex-field TRG to within $10$--$20\%$ for $0.05\le\beta\le 0.4$. The two shape parameters $(a,b)$ are fixed at each $\beta$ by the first two
sum rules in Eq.~\eqref{eq:sumrules}: Onsager's free energy and the exact strong-field coefficient $L_1 = e^{-8\beta}$. No parameter has been adjusted
to the TRG data; representative values are given in Table~\ref{tab:params}.

Two features of Eq.~\eqref{eq:New2} implement the locality requirement of (c). Below $T_c$, the weight $M_0$ of the infinitely-coupled chain is taken from the
region $v\lesssim v_1$, while for $v\gg v_1$ the distribution is unchanged, $B(a,b)^{-1}v^{a-1}(1-v)^{b-1}$; this also makes the zero-field susceptibility in the
ordered phase finite. Above $T_c$, the term $\omega(v_0/v)^n$ puts the weight removed from the gap $v<v_0$ back next to the Yang--Lee edge: with the value of
$\omega$ given above, the leading non-analytic correction $\propto v_0^{a}\propto t^{15a/4}$, which would otherwise multiply the whole distribution, cancels; the remaining
non-analytic terms are of order $v_0^{1+a}\propto t^{3.9}$. The edge behaviour $(v-v_0)^{-2/3}$, i.e.\ the exponent $-1/6$ of the Yang--Lee edge, is retained.
The exponent $n$ controls how sharply this weight is concentrated. Its value at $T_c$, $n=1$, is our choice (the simplest power for which the leftover terms are
smooth; the maximum deviation from TRG ranges from $0.07\%$ to $0.19\%$ for $0.5\le n(T_c)\le 2$), and the term $6v_0$, irrelevant near $T_c$, is needed for
a positive solution of the sum rules at the highest temperatures ($\beta\lesssim 0.1$); the results do not depend on its precise value. These are the only chosen numbers
in Eq.~\eqref{eq:New2}. As a consequence, at fixed $h\neq 0$ the free energy and its first derivative are continuous in $\beta$ across $\beta_c$ and the specific heat is
positive; since $(a,b)$ are fixed by Onsager's free energy, the specific heat inherits a weak logarithmic divergence at $\beta_c$ from Onsager's $t^2\ln|t|$
(with an amplitude reduced by a factor that decreases with $h$, e.g.\ $0.7$ at $h=0.1$), on top of a finite jump.\footnote{If, instead, the atom $M_0$ is simply
added to a beta distribution of weight $1-M_0$ and the edge term $\omega(v_0/v)^n$ is omitted above $T_c$, the maximum deviation from TRG is smaller
($0.08\%$, due to the high-temperature branch), but terms $\propto |t|^{1/8}$ and $t^{15a/4}$ then enter the free energy at fixed $h\neq0$, which leads to a divergent
internal energy and a negative specific heat close to $T_c$.}

By construction Eq.~\eqref{eq:New2} reduces to Onsager's result at $h=0$, gives Yang's spontaneous magnetization, obeys the Lee--Yang theorem,
and reproduces the strong-field expansion to order $z$; at low temperature $W_\beta$ tends to the uniform distribution of (e). On the critical isotherm,
$v_1=0$ and the magnetization is a Gauss hypergeometric function,
\begin{equation}
\label{eq:M2F1}
M_{\rm II}(\beta_c, h) = \pFq{2}{1}{\frac{1}{2},\, a_c}{a_c+b_c}{-\frac{1}{\sinh^2(\beta_c h)}};
\end{equation}
in general, $M_{\rm II}$ is given by the one-dimensional integral Eq.~\eqref{eq:Mchains} (an Appell function), and
$-\beta f_{\rm II}(\beta,h) = -\beta f(\beta,0) + \int_0^{\beta h} M_{\rm II}\, dH$.

\begin{table*}
\caption{\label{tab:params} Parameters of Conjecture~II, Eq.~\eqref{eq:New2}, fixed by the first two sum rules of Eq.~\eqref{eq:sumrules} (no fit), and the resulting predictions for the
Lee--Yang point $\beta h = i\pi/2$, which was not used as input [exact values from Eqs.~\eqref{eq:Merlini1} and \eqref{eq:LYM}].}
\begin{ruledtabular}
\begin{tabular}{ccccccc}
$\beta$ & $M_0$ or $\sqrt{v_0}$ & $a$ & $b$ & $v_1$ or $n$ & $-\beta{\rm Re} f(\beta,\frac{i\pi}{2\beta})$: Eq.~\eqref{eq:New2} / exact & $M(\beta, \frac{i\pi}{2\beta})$: Eq.~\eqref{eq:New2} / exact \\
\hline
0.20 & $\sqrt{v_0}=0.2753$ & 0.1265 & 0.9212 & $n=1.455$ & 0.2437 / 0.2567 & 1.264 / 1.179 \\
0.30 & $\sqrt{v_0}=0.0912$ & 0.1156 & 1.3584 & $n=1.050$ & 0.5346 / 0.5336 & 1.090 / 1.092 \\
0.40 & $\sqrt{v_0}=0.0079$ & 0.0604 & 1.4274 & $n=1.000$ & 0.7706 / 0.7684 & 1.040 / 1.048 \\
$\beta_c$ & $M_0 = 0$ & 0.03282 & 1.0823 & $v_1 = 0$ & 0.8578 / 0.8580 & 1.0375 / 1.0366 \\
0.45 & $M_0 = 0.7493$ & 0.02850 & 1.0142 & $v_1=3.9\times10^{-5}$ & 0.8775 / 0.8782 & 1.0377 / 1.0344 \\
0.50 & $M_0 = 0.9113$ & 0.01770 & 0.9217 & $v_1=6.0\times10^{-3}$ & 0.9841 / 0.9849 & 1.0293 / 1.0246 \\
0.60 & $M_0 = 0.9736$ & 0.00929 & 0.9204 & $v_1=6.8\times10^{-2}$ & 1.1926 / 1.1929 & 1.0140 / 1.0123 \\
1.00 & $M_0 = 0.9993$ & 0.00153 & 0.9911 & $v_1=1.67$ & 1.9997 / 1.9997 & 1.0006 / 1.0006 \\
\end{tabular}
\end{ruledtabular}
\end{table*}

\emph{Critical exponent $\delta$.} At $\beta=\beta_c$, Eq.~\eqref{eq:M2F1} gives, using the behaviour of $\,_2F_1$ at large negative argument,
\begin{equation}
\label{eq:delta}
M_{\rm II}(\beta_c, h) \simeq \frac{\Gamma(a_c+b_c)\,\Gamma(\frac{1}{2}-a_c)}{\sqrt{\pi}\,\Gamma(b_c)}\, (\beta h)^{2a_c}, \qquad \delta = \frac{1}{2a_c} \, ,
\end{equation}
where $(a_c, b_c)$ solve the two sum rules at $\beta_c$, whose right-hand sides are known in closed form,
$a_c/(a_c+b_c) = e^{-8\beta_c} = (\sqrt{2}-1)^4$ and $\langle \ln(1+\sqrt{v})\rangle = 2G/\pi + \frac{1}{2}\ln 2 - \ln(1+\sqrt{2}) = 0.0483218$ ($G$ is Catalan's constant).
We find $a_c = 0.0328249$, $b_c = 1.082256$, hence
\begin{equation}
\delta = 15.23, \qquad M_{\rm II}(\beta_c,h) \simeq 1.0547\, (\beta h)^{1/15.23},
\end{equation}
to be compared with the exact $\delta=15$ and amplitude $1.0590$ of Eq.~\eqref{eq:isotherm}. Thus, using as input only Onsager's
zero-field free energy and the energy of a single flipped spin, the improved expression \emph{predicts} the critical isotherm exponent to
$1.5\%$ and its amplitude to $0.4\%$; since the exponents differ slightly, $M_{\rm II}(\beta_c,h)$ deviates from the exact isotherm by $0.3$--$2\%$ for
$10^{-9}<\beta h<10^{-3}$. This prediction is robust: fixing $(a_c,b_c)$ instead by other pairs of the exact sum rules in Eq.~\eqref{eq:sumrules}
gives $14.9\le \delta \le 15.2$ and amplitudes between $1.0547$ and $1.0569$ in eight of the ten possible cases (Table~\ref{tab:delta}); in particular, the Lee--Yang data
alone give $\delta=14.94$. Only the pairs that combine the second moment with a Lee--Yang rule behave differently ($\delta=14.60$ for the root closest to the others,
a second root has $\delta=2.4$; or no solution).
The beta distribution with the parameters of Eq.~\eqref{eq:New2} satisfies all five rules at $\beta_c$ to better than $1\%$, and the Lee--Yang magnetization, which is not an input,
is reproduced to $8\times 10^{-4}$ (Table~\ref{tab:params}). The zero-field susceptibility diverges with the exponents $(15/8)(1-2a_c)=1.752$ above and
$(1-2a_c)/(16a_c) = 1.78$ below $T_c$, close to $\gamma = 7/4$; its amplitude is about $1.4$ times the exact $C_0^+$ above $T_c$, and about a third of
the exact $C_0^-$ just below $T_c$.

\begin{table}
\caption{\label{tab:delta} Critical-isotherm exponent and amplitude, $M(\beta_c,h)\simeq A_M\,(\beta h)^{1/\delta}$, obtained from the beta distribution at
$\beta_c$ when its two parameters are fixed by different pairs of the exact sum rules of Eq.~\eqref{eq:sumrules}; the first row is Conjecture~II, Eq.~\eqref{eq:New2}. ``On'' denotes Onsager, ``$L_1$''
and ``$L_2$'' the first and second moment rules, and ``LY-$f$'' and ``LY-$M$'' the free energy and magnetization at $\beta h=i\pi/2$.}
\begin{ruledtabular}
\begin{tabular}{lcc}
sum rules & $\delta$ & $A_M$ \\
\hline
On, $L_1$ [Eq.~\eqref{eq:New2}] & 15.23 & 1.0547 \\
On, $L_2$ & 15.21 & 1.0549 \\
On, LY-$f$ & 15.15 & 1.0555 \\
On, LY-$M$ & 15.11 & 1.0559 \\
$L_1$, $L_2$ & 15.15 & 1.0553 \\
$L_1$, LY-$f$ & 14.94 & 1.0569 \\
$L_1$, LY-$M$ & 14.94 & 1.0569 \\
LY-$f$, LY-$M$ & 14.94 & 1.0569 \\
$L_2$, LY-$M$ & 14.60 & 1.0589 \\
$L_2$, LY-$f$ & \multicolumn{2}{c}{no solution} \\
\hline
exact & 15 & 1.0590 \\
\end{tabular}
\end{ruledtabular}
\end{table}

\section{Comparison with TRG}
\label{sec:compare}
\begin{figure}[tbp]
	\centering
	\includegraphics[width=0.45\textwidth]{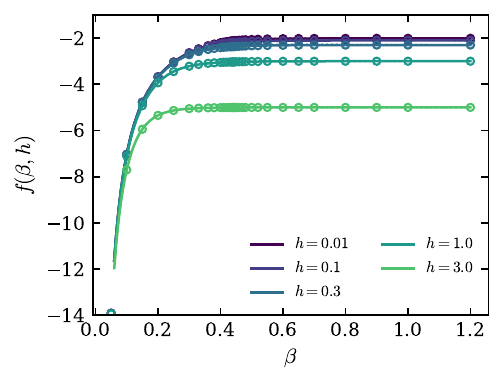}
	\caption{\label{fig:2}The free energy $f(\beta,h)$ from Conjecture~II, Eq.~\eqref{eq:New2} (solid lines), and from Conjecture~I,
	Eq.~\eqref{eq:New1} (dotted lines), compared with HOTRG (hollow markers). On this scale Conjecture~II$'$ (Eqs.~\eqref{eq:refined_f}--\eqref{eq:refined_edge}) is indistinguishable from Conjecture~II.}
\end{figure}

\begin{figure}[tbp]
	\centering
	\includegraphics[width=0.45\textwidth]{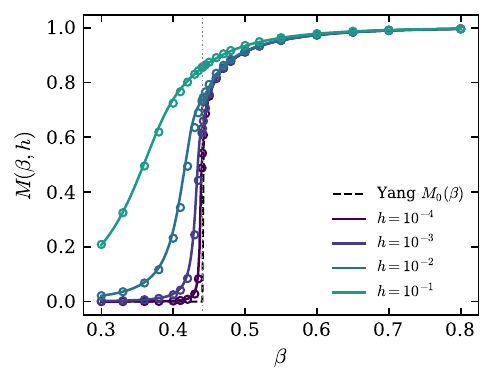}
	\caption{\label{fig:mag}Magnetization from Conjecture~II, Eqs.~\eqref{eq:New2} and \eqref{eq:Mchains} (lines), and from HOTRG (markers). The dashed line is Yang's spontaneous
	magnetization, Eq.~\eqref{eq:Yang}. Conjecture~I, Eq.~\eqref{eq:New1}, gives $M\to 0$ as $h\to 0$ at all temperatures.}
\end{figure}

The free energy obtained from Conjecture~II, Eq.~\eqref{eq:New2}, is compared with TRG in Fig.~\ref{fig:2}, and the relative deviations of Conjectures~I, II and II$'$
[Eqs.~\eqref{eq:New1}, \eqref{eq:New2} and Eqs.~\eqref{eq:refined_f}--\eqref{eq:refined_edge}] are shown in the three panels of Fig.~\ref{fig:error}. The maximum relative deviation over our grid ($0.05\le \beta \le 2$, $0 < h \le 5$) is $0.15\%$ (at $\beta = 0.40$, $h=0.2$),
compared with $2.6\%$ for Conjecture~I, Eq.~\eqref{eq:New1}; the root-mean-square deviations are $0.03\%$ and $0.9\%$, respectively. Conjecture~II$'$
reduces the maximum deviation to $4\times10^{-5}$ (Table~\ref{tab:errors}). In the ordered phase the
deviation is at most $2.0\times10^{-4}$ and decreases rapidly with $\beta$ ($<10^{-5}$ for $\beta>0.6$); see Table~\ref{tab:errors}. The largest deviations occur just above $T_c$,
where the susceptibility of Eq.~\eqref{eq:New2} is too large (by $7$--$17\%$ at $\beta=0.38$--$0.41$, and by up to $40\%$ as $T\to T_c^+$).
The magnetization of Conjecture~II, Eq.~\eqref{eq:Mchains}, is shown in Fig.~\ref{fig:mag}: it has the correct spontaneous magnetization and describes the TRG data well away from
$T_c$ (to about $10^{-2}$ or better for $\beta\ge0.445$); close to $T_c$, in the crossover between critical and off-critical behaviour at small $h$, the deviations reach $0.05$--$0.09$
(the largest, $0.09$, at $\beta=0.435$, $h=10^{-3}$); for Conjecture~II$'$, Eq.~\eqref{eq:refined_M}, the deviations are below $8\times10^{-4}$ for $h\geq10^{-4}$.
The critical isotherm of all three conjectures is shown in Fig.~\ref{fig:isotherm}.
Finally, Fig.~\ref{fig:LY} compares the Lee--Yang zero density implied by Conjecture~II [Eq.~\eqref{eq:New2} inserted in Eq.~\eqref{eq:Mchains} at complex field] with the one computed with TRG.

\begin{table}
\caption{\label{tab:errors} Maximum relative deviation $|f_{\rm conj}/f_{\rm TRG}-1|$ over $0<h\le5$ in different temperature ranges for the original
conjecture (I, Eq.~\eqref{eq:New1}), the improved conjecture (II, Eq.~\eqref{eq:New2}) and the refined form of Appendix~\ref{app:refined} (II$'$, Eqs.~\eqref{eq:refined_f}--\eqref{eq:refined_edge}). The last row gives the maximum and, in parentheses,
the root-mean-square deviation over the whole grid.}
\begin{ruledtabular}
\begin{tabular}{lccc}
range of $\beta$ & I & II & II$'$ \\
\hline
$[0.05, 0.25]$ & $2.7\times10^{-3}$ & $6.2\times10^{-4}$ & $2.1\times10^{-5}$ \\
$(0.25, \beta_c)$ & $2.4\times10^{-2}$ & $1.5\times10^{-3}$ & $4.1\times10^{-5}$ \\
$[\beta_c, 0.6]$ & $2.6\times10^{-2}$ & $2.0\times10^{-4}$ & $2.9\times10^{-5}$ \\
$(0.6, 2]$ & $2.0\times10^{-2}$ & $3.1\times10^{-6}$ & $6.9\times10^{-7}$ \\
\hline
all & $2.6\times10^{-2}$ & $1.5\times10^{-3}$ & $4.1\times10^{-5}$ \\
    & ($8.8\times10^{-3}$) & ($2.7\times10^{-4}$) & ($7.3\times10^{-6}$) \\
\end{tabular}
\end{ruledtabular}
\end{table}

\section{Summary}
We revisited our conjecture for the free energy of the two-dimensional Ising model in a real magnetic field. Precise HOTRG data show that the
original expression, Conjecture~I [Eq.~\eqref{eq:New1}], is accurate to $2.6\%$ but misses the spontaneous magnetization and the critical behaviour
($\delta_{\rm eff}\to 1$). Using the Lee--Yang theorem we showed that the exact free energy is a superposition of one-dimensional Ising chains in
the same field, Eq.~\eqref{eq:chains}, of which Eq.~\eqref{eq:New1} is a single-chain approximation, and translated the known exact results into
sum rules for the chain distribution, Eq.~\eqref{eq:sumrules}. The improved conjecture, Conjecture~II [Eq.~\eqref{eq:New2}], uses as input only Onsager's free energy, Yang's spontaneous magnetization,
the energy of a single flipped spin and, above $T_c$, the location of the Yang--Lee edge in the scaling limit.
It agrees with TRG to $0.15\%$ for all $\beta$ and $h$ (rms $0.03\%$, thirty times better than the original conjecture), gives the spontaneous magnetization exactly,
satisfies the Lee--Yang theorem, has a positive specific heat, and predicts the critical-isotherm exponent $\delta = 15.2$ and amplitude $1.055$, within
$1.5\%$ and $0.4\%$ of the exact values; most other choices of exact sum rules give $14.9\le\delta\le15.2$.

The expression is not exact. Its known shortcomings are: the predicted $\delta$ differs from $15$; the susceptibility amplitudes are off (by a factor of about
$1.4$ above and $1/3$ below $T_c$), so that the crossover of the magnetization near $T_c$ at small $h$ is only approximately reproduced; at fixed $h\neq0$ the
specific heat is not analytic at $\beta_c$ (a finite jump and a weak logarithmic divergence inherited from Onsager's solution); and at high temperature ($\beta\lesssim0.25$),
where the exact chain distribution is not positive, the predicted Lee--Yang density near $\theta=\pi$ is too large. The representation Eq.~\eqref{eq:chains} is, however,
systematically improvable. As shown in Appendix~\ref{app:refined} [Conjecture~II$'$, Eqs.~\eqref{eq:refined_f}--\eqref{eq:refined_edge}], imposing more of the exact sum rules together with the scaling amplitudes of the Ising
field theory, plus a few constants adjusted to transfer-matrix and TRG data, gives a refined form with $\delta=15$ and the exact amplitudes by construction,
which agrees with TRG to $4\times 10^{-5}$ in the free energy and $8\times 10^{-4}$ in the magnetization (for $h\geq10^{-4}$). We believe this representation is a useful framework
for the problem: the exact solution, if it is ever found, will be a particular distribution of one-dimensional Ising chains.

\subsection*{Acknowledgements}
The first version of this manuscript was prepared by a human. However, it was not accurate enough to reproduce the critical exponent $\delta$. The original conjecture in the manuscript refers to this version.
The second version (with improvement in conjecture) was prepared with a large language model. In October 2026, working in a new world, I asked Claude Code Opus 5.5 to attack this problem and this led to the improved conjecture(s) that are far more accurate. I have checked the results and take full responsibility for the content. After first version, I heard back from Jacques H.H. Perk, Gandhi Viswanathan, Vikram Vyas, and Debayan Chakraborty and would like to thank them for finding this work interesting. This material is based upon work supported by the U.S. Department of Energy, Office of Science, Office of Nuclear Physics under contract DE-AC05-06OR23177 and the U.S. Department of Energy, Office of Science, National Quantum Information Science Research Centers, Co-design Center for Quantum Advantage under contract number DE-SC0012704. 

\subsection*{Data Availability Statement}

The TRG notebook and the data for the free energy to compare to original conjecture can be found
\href{https://zenodo.org/records/15285090}{on Zenodo}. The data and codes used in this version can be requested from the author.

\bibliographystyle{utphys}
%%%%%%%%%%%%%%%%%%%%%%%%%%%%%%%%%%%%%%%
\raggedright
\bibliography{v2.bib}

\appendix
\section{More refinement -- Conjecture II\texorpdfstring{$'$}{'}}
\label{app:refined}
The representation Eq.~\eqref{eq:LYrep} can be used to build in more information. \textbf{We call the result Conjecture~II$'$}; it is completely defined by
Eqs.~\eqref{eq:refined_f}--\eqref{eq:refined_edge} below. Write $m(\theta)\equiv 2\pi\rho(\theta)$ for the Lee--Yang zero density, $\varsigma = \sin\frac{\theta}{2}$, $k = \sinh^{-2}(2\beta)$ and $u=e^{-4\beta}$.
The free energy and magnetization follow from Eq.~\eqref{eq:LYrep}:
\begin{align}
-\beta f_{\rm II'}(\beta,h) &= -\beta f(\beta,0) \nonumber \\ &+ \frac{1}{2\pi}\int_{\theta_0}^{\pi} m(\theta)\,\ln\Big(1+\frac{\sinh^2 \beta h}{\varsigma^2}\Big)\,d\theta , \label{eq:refined_f} \\
M_{\rm II'}(\beta,h) &= \frac{1}{\pi}\int_{\theta_0}^{\pi} m(\theta)\,\frac{\sinh 2\beta h}{\cosh 2\beta h-\cos\theta}\,d\theta , \label{eq:refined_M}
\end{align}
where $-\beta f(\beta,0)$ is Onsager's result, Eq.~\eqref{eq:HPQ1}. The zero density is taken to be
\begin{equation}
\label{eq:refined_m}
m(\theta) = \begin{cases} \mathcal{W}(\varsigma;\beta)\, P_\beta(\cos\theta), & \theta_0<\theta\le\pi, \\ 0, & 0\le\theta<\theta_0, \end{cases}
\end{equation}
with $\theta_0 = 0$ for $\beta\ge\beta_c$ and $\theta_0$ the Yang--Lee edge, Eq.~\eqref{eq:refined_edge}, for $\beta<\beta_c$. The factor $\mathcal{W}$, Eq.~\eqref{eq:refined_W},
carries the singular behaviour, and $P_\beta$ is a polynomial of degree six in $\cos\theta$ whose seven coefficients are fixed by the seven linear conditions
(all integrals over $\theta_0<\theta<\pi$)
\begin{align}
&\frac{1}{\pi}\int m\,d\theta = 1, \qquad P_\beta(\cos\theta_0)=1, \nonumber \\
&\frac{1}{\pi}\int m\ln(2\varsigma)\,d\theta = -\beta f(\beta,0)-2\beta, \nonumber \\
&\frac{1}{\pi}\int m\ln\Big(2\cos\frac{\theta}{2}\Big)\,d\theta = -\beta\,{\rm Re}\,f\Big(\beta,\frac{i\pi}{2\beta}\Big) - 2\beta, \nonumber \\
&\frac{2}{\pi}\int m\cos(n\theta)\,d\theta = -2nL_n(u) \quad (n=1,2,3). \label{eq:refined_cond}
\end{align}
Apart from the convention $P_\beta(\cos\theta_0)=1$, these are the normalization and the sum rules of Eq.~\eqref{eq:sumrules} [with Eqs.~\eqref{eq:HPQ1} and \eqref{eq:Merlini1}
on the right-hand side] written for $m$; the strong-field coefficients $L_1$, $L_2$ are given below Eq.~\eqref{eq:strong} and $L_3 = 6u^4 - 16u^5 + \frac{31}{3}u^6$.
The singular factor is
\begin{widetext}
\begin{align}
\mathcal{W} &= M_0\Big[(1-r)+r\Big((1+y^2)^{4/15}+e_1\,\frac{y^2}{(1+y^2)^{3/2}}\Big)\Big]^{1/8}, \quad y = \frac{\varsigma}{\alpha}, \quad \alpha=\Big(\frac{rM_0^8}{\mathcal{B}^8k}\Big)^{15/8} \qquad (\beta>\beta_c), \nonumber \\
\mathcal{W} &= \mathcal{B}\,\varsigma^{1/15} \qquad (\beta = \beta_c), \nonumber \\
\mathcal{W} &= \mathcal{B}\,k^{1/6}\varsigma^{1/15}\Big(\frac{\varsigma^2}{\varsigma^2-\varsigma_0^2}\Big)^{1/6}\Big(1-\lambda_0\zeta^{-8/15}-\lambda_2\zeta^{-2}\Big), \quad \zeta = \frac{\varsigma}{\varsigma_0}, \quad
\varsigma_0 = \sin\frac{\theta_0}{2} \qquad (\beta<\beta_c), \label{eq:refined_W}
\end{align}
\end{widetext}
with $M_0$ from Eq.~\eqref{eq:Yang} and $\mathcal{B} = 1.058986\cos\frac{\pi}{30}$, so that the critical isotherm Eq.~\eqref{eq:isotherm} is reproduced exactly. The Yang--Lee edge above $T_c$ is
\begin{equation}
\label{eq:refined_edge}
\tan\frac{\theta_0}{2}=\frac{\xi_0}{s}\,\frac{\eta^{15/8}}{(1+b_1\eta+b_2\eta^2)^{7/16}},
\end{equation}
with $\eta=(1-q)/\sqrt{q}$ and $q = e^{2\beta}\tanh\beta$,
which reduces to the scaling-limit edge of Sec.~\ref{sec:exact} as $T\to T_c^+$.
The constants in Eqs.~\eqref{eq:refined_W} and \eqref{eq:refined_edge} are as follows. With the linear-in-$t$ coefficient of the Ising field theory free energy at $T_c$, $\Phi_1 = -0.3188096$~\cite{FLZZ1998, FonsecaZamolodchikov2003},
and $\Phi_0 = 1.1977334$, one has $c_1 = \frac{|\Phi_1|}{2\Phi_0}\frac{\cos(7\pi/30)}{\cos(\pi/30)} = 0.099449$, $\lambda_0 = c_1\xi_0^{-8/15} = 0.24159$ and
$r = 1-2\sqrt{2}\,c_1\mathcal{B}^8 s^{-8/15} = 0.53234$. The constants $\lambda_2 = 0.086543$ and $e_1 = 0.45132$ are fixed by the exact susceptibility amplitudes $C_0^{+}$ and $C_0^{-}$~\cite{WMTB1976, Orrick2001}.
Finally, $b_1 = 0.2667$ and $b_2 = (w_c s^2/\xi_0^2)^{-8/7} = 0.4908$ with $w_c = 0.095$, as well as the exponents $\frac16$ of $k^{1/6}$ and $\frac32$ in the low-temperature
factor, were chosen by comparison with Yang--Lee edges from transfer matrices and with the TRG data. These chosen constants matter: without $b_1$ and $b_2$ the
maximum deviation from TRG is $1.7\times10^{-3}$. \textbf{Conjecture~II$'$ has $\delta = 15$ and the exact critical amplitude}, the exact susceptibility amplitudes above and below $T_c$, the Yang--Lee edge exponent $-1/6$,
Yang's $M_0$, and the strong-field expansion through $O(z^3)$; its zero density is positive for $0.02\le\beta\le3$. The price is a less transparent, piecewise construction
with a few chosen constants, in which $P_\beta$ is defined implicitly by the linear system Eq.~\eqref{eq:refined_cond}; moreover, at fixed $h\neq0$ the free energy has a kink in $\beta$ at $\beta_c$.
Its deviation from TRG is shown in the bottom panel of Fig.~\ref{fig:error} and in Table~\ref{tab:errors}, and its critical isotherm in Fig.~\ref{fig:isotherm}.

\end{document}